\documentclass[%
 aip,
 amsmath,amssymb,
 reprint,%
]{revtex4-1}

\usepackage{graphicx}
\usepackage{dcolumn}
\usepackage{bm}

\usepackage[utf8]{inputenc}
\usepackage[T1]{fontenc}
\usepackage{mathptmx}

\usepackage{color,amssymb}
\definecolor{orange}{rgb}{1,0.5,0} 

\begin{document}


\title[Time-resolved rheometry of drying liquids and suspensions]{Time-resolved rheometry of drying liquids and suspensions}

\author{Pierre Leh\'ericey}
 \affiliation{Department of Materials Science and Engineering, Massachusetts Institute of Technology, Cambridge, MA 02139, USA\looseness=-1}

\author{Audrey Delots}
 \affiliation{Department of Materials Science and Engineering, Massachusetts Institute of Technology, Cambridge, MA 02139, USA\looseness=-1}

\author{Niels Holten-Andersen}%
\affiliation{Department of Materials Science and Engineering, Massachusetts Institute of Technology, Cambridge, MA 02139, USA\looseness=-1}%

\author{Thibaut Divoux}
\email{Thibaut.Divoux@ens-lyon.fr}
\affiliation{Univ Lyon, Ens de Lyon, Univ Claude Bernard, CNRS, Laboratoire de Physique, F-69342 Lyon, France\looseness=-1}%
\affiliation{MultiScale Material Science for Energy and Environment, UMI 3466, CNRS-MIT, 77 Massachusetts Avenue, Cambridge, MA 02139, USA\looseness=-1}

\date{\today}

\begin{abstract}
From paints and inks to personal care and food products, solvent evaporation is ubiquitous and critically impacts product rheological properties. It affects Newtonian fluids, by concentrating any non-volatile components, as well as viscoelastic materials such as gels, which hardens up and become solid upon complete drying. In both of these cases, solvent evaporation also leads to a change in volume of the sample, which makes any measurements of resulting rheological properties particularly challenging with traditional shear geometries. Here we show that the rheological properties of a sample experiencing `slow' evaporation can be monitored in a time-resolved fashion by using a zero normal-force controlled protocol in a parallel-plate geometry. In this framework, the solvent evaporation from the sample leads to a decrease of the normal force, which is compensated at all times by a decrease of the gap height between the plates. As a result, the sample maintains a constant contact area with the plates despite the significant decrease of its volume. We validate the method under both oscillatory and continuous shear by accurately monitoring the dynamic and shear viscosities of samples of water and water-glycerol mixtures experiencing evaporation for several hours, and a corresponding relative volume decrease as large as 70\%. Moreover, we show that this protocol is also applicable for characterizing complex fluids such as suspensions experiencing evaporation. Specifically, we monitor a colloidal dispersion of charged silica nanoparticles undergoing a glass transition induced by evaporation. While the decrease in gap height between the plates provides a direct estimate of the increasing particle volume fraction, oscillatory and continuous shear measurements allow us to monitor the evolving viscoelastic properties of the suspension in real time. Overall, our study shows that a zero normal-force protocol can provide a simple approach to bulk and time-resolved rheological characterization of a variety of simple and complex fluid systems experiencing slow volume variations.      
\end{abstract}

\maketitle

\section{Introduction}
    Evaporation of volatile liquids alter the rheological properties of numerous materials from Newtonian liquids (e.g., water in water-glycerol mixtures) to viscoelastic liquids and soft solids (e.g., volatile solvents in suspensions and gels) \cite{Boger:1969,Macosko:1994}. On the one hand, evaporation can be an issue that needs to be limited, for instance when performing rheological measurements to characterize functional properties of samples with volatile solvents, such as paints or inks, etc. \cite{Ewoldt:2015}. In that case, rheological experiments are performed with a solvent trap, or an oil rim at the sample periphery for parallel-plate and cone-and-plate geometry, or even in an insulated chamber \cite{Orafidiya:1989,Wee:1998,Sato:2005}. On the other hand, evaporation can be used to alter the microstructure of a sample and tune its properties, often through a sol-gel or sol-glass transition. For instance, solvent evaporation plays a key role in the formation of films with self-assembled nanostructure \cite{Lee:2006,Prevo:2007}. Moreover, solvent evaporation can be used to concentrate a suspension of particles into a solid, whose microstructure and mechanical properties can be accurately controlled  \cite{Leng:2006,Merlin:2012,Lidon:2014,Piroird:2016}. Similarly, a solvent can also be evaporated from a polymer mixture to create a hardened composite. Contrary to UV-curing where the addition of a photo-initiator is needed to form an interconnected network from low molecular weight components through photo-polymerization \cite{Endruweit:2006}, solvent evaporation can happen with or without curing agents. This is the case of coatings  where the formation of intermolecular or interparticle bonds start after evaporation \cite{Poth:2001, Zhou:2013}. Finally, natural materials such as skin, hair and horns are classical examples of extra-organismal biomacromolecular materials that undergo various levels of dehydration in time and space as part of their life cycle \cite{Fudge:2009}. In fact, material dehydration has been documented as an effective biological material design strategy for achieving unique mechanical properties, as in some cases the synergistic effect of dehydration and local macromolecular network densification can lead to exceptional mechanical strength. This is the case of the beak of the Humboldt Squid, which shows a strong gradient in water content, and whose tip  is among the hardest and stiffest wholly organic materials known \cite{Miserez:2008}. 
    
    A key challenge in studies of complex fluids or soft solids with volatile solvents remains how to quantify the changes induced by solvent evaporation on the rheological properties of a system experiencing dehydration. Evaporation is a non-isochoric process, which often involves spatially heterogeneous dynamics that leads to concentration gradients \cite{Zakharov:2010}. As a result, time-resolved rheological measurements at the macroscale are very challenging \cite{Molenaar:1997}. This is the reason why most of the experimental effort has been focused on indirect measurements, e.g., based on film  surface  topography \cite{Figliuzzi:2012}, and local investigation techniques such as active and passive micro-rheology \cite{Bodiguel:2010,Kim:2013,Komoda:2014,Castro:2017}.   
    
     Here we propose to perform time-resolved measurements of the macroscopic rheological properties of samples experiencing slow evaporation, i.e., slow enough to safely assume that the sample does not develop any spatial plug stopping dehydration. Specifically, after placing the sample in a parallel-plate geometry connected to a stress-controlled rheometer, we use a zero normal-force (ZNF) protocol, in which the normal force exerted on the sample is maintained equal to zero, while the gap height is free to adjust. This method, which was previously shown to be efficient in measuring rheological properties during non-isochoric gelation in agarose dispersions \cite{Mao:2016,Mao:2017} and organogels \cite{Beniazza:2020}, is employed here on Newtonian fluids and charged colloidal suspensions. We show that the ZNF protocol allows us to monitor the continuous increase in viscosity of a dehydrating water-glycerol mixture, as well as the discontinuous changes in viscoelastic properties of a drying colloidal suspension during its glass transition. This study thereby paves the way for systematic time-resolved rheological measurements on macroscopic samples of complex fluids experiencing strongly non-isochoric transitions induced by solvent evaporation. 

     The article is organized as follows: in section~\ref{MatandMet}, we describe the samples characterized in the drying experiments and motivate the ZNF protocol. In section~\ref{Results}, we validate this protocol by confirming the invariant viscosity of a drying water sample monitored over several hours. We then apply the ZNF protocol to monitor the increasing viscosity of drying water-glycerol mixtures as water progressively evaporates. Drying experiments performed either under oscillatory or continuous shear demonstrate that the ZNF protocol allows us to monitor sample viscosity with high precision, while the sample volume decreases by up to 70\%. Finally, we apply the ZNF protocol to monitor the drying of a stable colloidal suspension. Here, we further validate the ZNF protocol by confirming that the estimate of the critical volume fraction associated with the dehydration-induced glass transition is in good agreement with values from the literature. Moreover, interpretations based on image correlation analysis from simultaneous video recordings collected during rheological measurements on the colloidal suspensions suggest that data collected under continuous shear, rather than oscillatory shear, provide more accurate information on dehydration-induced microstructural changes, likely due to suppress concentration gradients.

\section{Material and methods}
\label{MatandMet}
    \subsection{Rheological setup}
    \label{Rheosetup}
    Rheological measurements are performed with a parallel-plate geometry (diameter 25~mm), connected to a stress-controlled rheometer (MCR 302, Anton Paar). The bottom plate consists of a Peltier module (P-PTD 200), which sets the temperature of the sample. All the experiments were performed at $T=25^{\circ}$C. The upper plate is smooth and made of stainless steel (PP 25). Samples are loaded with a micropipette, which allows us to precisely fill the cell, whose initial gap is set by default to $h_0=500$~$\mu$m. The normal force is reset to $F_{\rm N}=0$~N immediately after loading the sample and prior to the start of the experiment, which simply consists of initiating the dehydration of the sample at constant temperature. The parallel-plate geometry is topped by a Peltier temperature-controlled hood (H-PTD 200), which consists of a cylinder (diameter 95~mm) centered with the rotation axis of the upper plate. The hood strongly reduces convection around the sample and allows us to control the evaporation rate by controlling the spacing between the foot of the hood and the bottom plate (the hood is typically placed at about 1~mm above the bottom plate). 
    In some experiments, the bottom plate and the Peltier module are replaced by a transparent glass plate to visualize the sample. In that case, images of the sample are taken through the transparent plate, which sits atop a support containing a semi-reflective blade forming a 45$^{\circ}$ angle with respect to the rotation axis of the rotor. The support is pierced with two 25~mm diameter holes; the first one is located under the sample, and the second one is facing the camera objective (Canon EF 100~mm, aperture $\digamma=2.8$ mounted on a DSLR Canon 6D), which is focused on the semi-reflective blade. In this configuration the entire surface of the plate is imaged with a spatial resolution of 352~$\mu$m/pixel.
    
\subsection{Samples}
In the present study, we monitor the drying of three different samples: ($i$) distilled water, ($ii$) water-glycerol mixtures with an initial weight fraction in glycerol (Sigma Aldrich, G9012, $\ge$ 99.5$\%$) ranging between 20\% wt. and 60\% wt., and ($iii$) a suspension of silica nanoparticles (Ludox HS-40, Sigma Aldrich) with a particle diameter  of 12~nm and a specific surface area of 220 m$^2$.g$^{-1}$.  The mass fraction of the as-purchased-dispersion is 40\% wt., which corresponds to an initial volume fraction of $\phi_0=0.23$. The suspension was degassed for 10~h under vacuum before testing. The colloids are stabilized by the presence of negative silanol groups at their free surface, which attract Na$^+$ cations in solution, hence forming an electrical double layer (EDL) responsible for the strong repulsive interaction between colloids. Several studies have shown that these colloids and their EDL display an elastic behavior well below the usual random close packing fraction. For Ludox HS-40, it was found that the critical volume fraction $\phi_c^{\rm (ref)}=0.32$ beyond which a solid-like response dominates \cite{Loussert:2016, DiGiuseppe:2012, Boulogne:2014, Ziane:2015}. 
\begin{figure}[t]
        \includegraphics[scale=0.25]{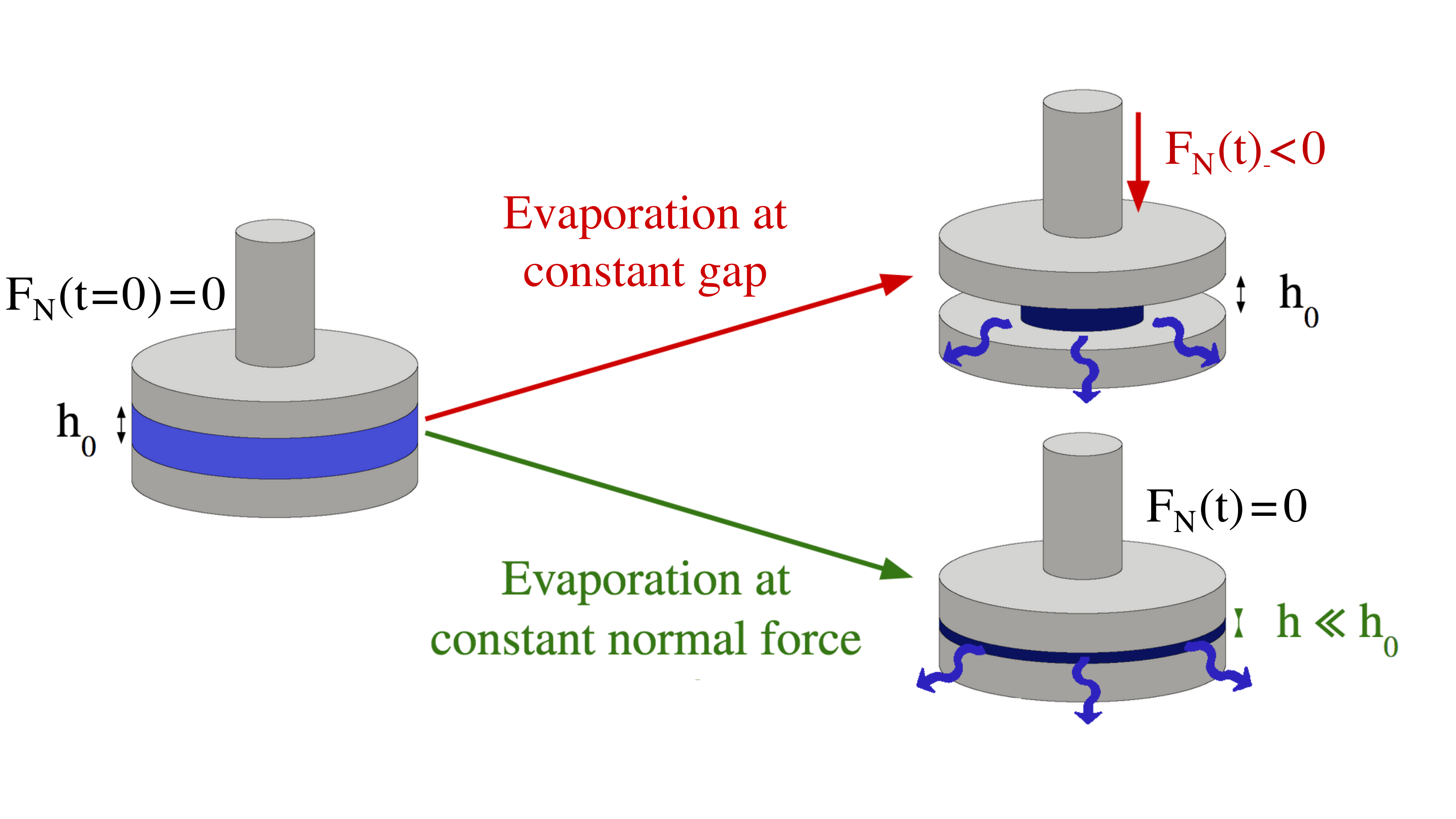}
        \caption{\label{fig1} Schematics of the comparison between drying experiments conducted at constant gap and constant normal-force, respectively. At constant gap, the evaporation leads to a decrease of the contact area between the sample and the plates. At constant normal-force, to maintain a zero normal force on the upper plate, the gap decrease compensates for the sample evaporation such that the apparent contact area between the plates and the sample remains the same.}
    \end{figure}

\subsection{Constant gap vs. zero normal force protocol}
Following the loading of the sample, the top plate is put in contact with the sample, and the normal force is reset to $F_{\rm N}=0$~N. As the volatile components of the sample slowly evaporates at the periphery of the plate, the volume of the sample decreases (Fig.~\ref{fig1}). 
On the one hand, if the gap height $h_0$ is maintained constant, the evaporation leads to a decrease of the curvature radius of the fluid meniscus at the periphery the cell. As a result, capillary forces pull on the upper plate of the geometry, leading to an increase of the absolute value of the normal force, which is about $|\Delta F_{\rm N}|=\gamma r_c$, where $\gamma$ is the surface tension at the air-fluid interface, and $r_c$ is the radius of curvature of the air/sample meniscus, with $r_c\simeq h_0/2$. In the first case, the sample considered is water and $|\Delta F_{\rm N}| \simeq 0.01$~N, which is larger than the sensitivity of the rheometer's normal force sensor (0.5~mN). Eventually, as the evaporation goes on, the contact line of the sample at the plate periphery contracts radially and the amount of sample left inside the geometry does not cover the entire surface of the plates, yielding a systematic error in the determination of the shear stress.   
On the other hand, if the normal force is maintained constant, equal to zero, while the gap height is left free to vary, the latter will decrease to compensate for the increase in the capillary forces induced by the sample evaporation, and indeed maintain $F_{\rm N}=0$~N. As a result, the sample will occupy a constant surface area identical to that of the plates, thus allowing for bulk rheological measurements, while the sample is slowly evaporating. In the rest of the paper, we validate and use this ZNF protocol to monitor the temporal evolution of the rheological properties of drying Newtonian fluids and viscoelastic samples.

\section{Results and discussion}
\label{Results}

\subsection{Measuring the viscosity of a drying Newtonian liquid}
\label{water}
\begin{figure}[t]
\includegraphics[scale=0.40]{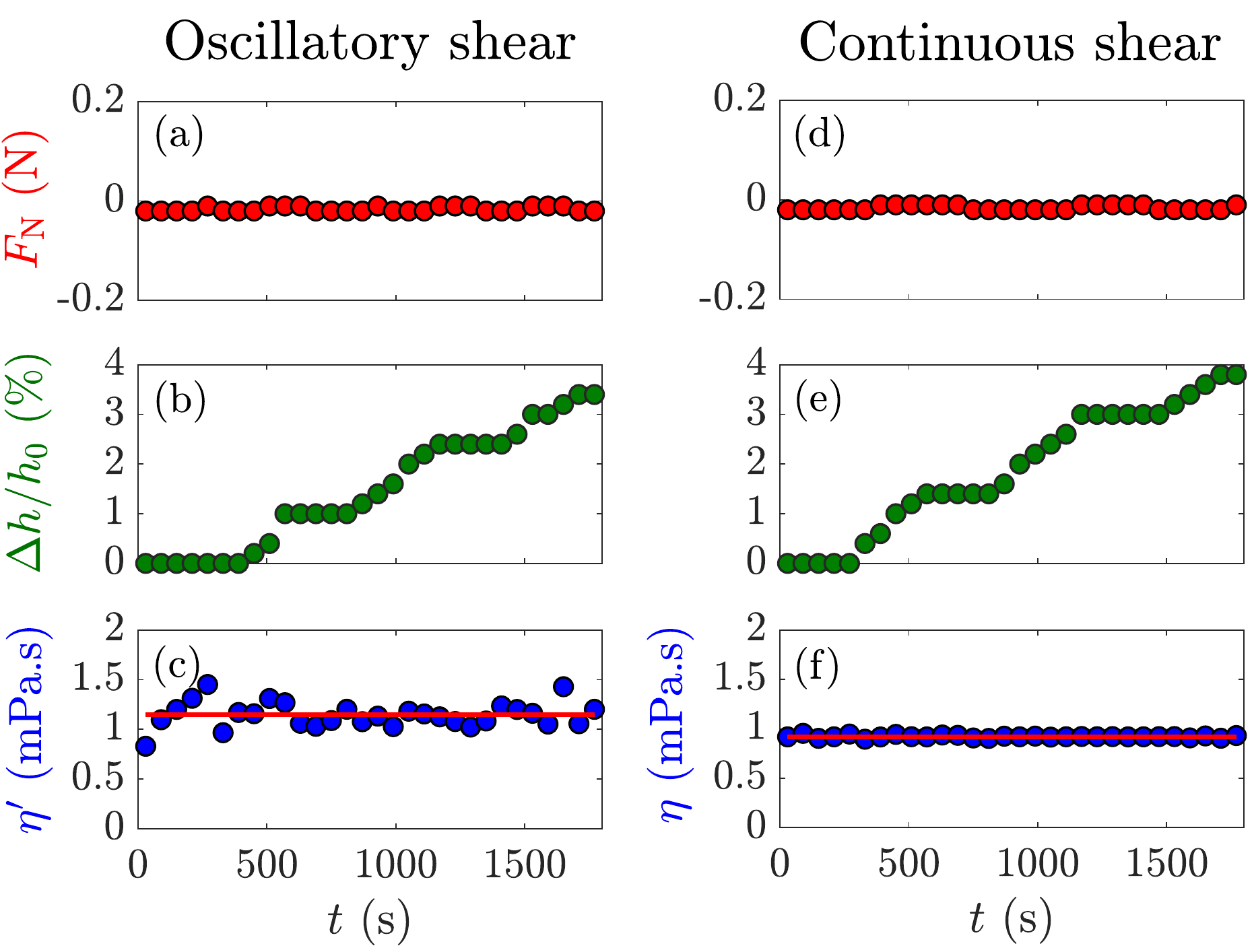}
\caption{\label{fig2} Viscosity measurement with the zero normal-force protocol under (a--c) oscillatory shear and (d--f) continuous shear on a drying water sample. In both cases, we report the temporal evolution of (a,d) the applied normal force $F_{\rm N}$, (b,e) the relative gap decrease $\Delta h/h_0$, (c) the dynamic viscosity $\eta'=G''/\omega_0$, and (f) the shear viscosity $\eta=\sigma / \dot{\gamma}$. Experiments performed at 25$^{\circ}$C, with $h_0=500$~$\mu$m. Oscillatory  measurements reported in (a--c) are performed at a frequency $f_0=\omega_0/(2\pi)=$0.2~Hz with a constant strain amplitude $\gamma_0=50\%$. Continuous shear measurements reported in (d--f) are performed at $\dot{\gamma}=$10 s$^{-1}$. In (c) and (f), the red line corresponds to the mean value of the viscosity computed over the duration of the experiment, i.e., $\eta'=(1.14\pm 0.12)$~mPa.s and $\eta=(0.92\pm 0.01)$~mPa.s, respectively. }
\end{figure}

To validate the ZNF protocol, we monitor the viscosity of a single-component Newtonian fluid sample composed of distilled water during evaporation from the parallel-plate geometry at $T=25^{\circ}$C over 30~min. The experiment is repeated twice in two different configurations. In the first experiment, the sample dynamics viscosity $\eta'=G''/\omega_0$, where $G''$ is the viscous modulus is assessed by oscillations of constant strain amplitude $\gamma_0$=50\% at a frequency $f_0=\omega_0/(2\pi)=0.2$~Hz. In the second experiment, the \textit{shear viscosity} $\eta=\sigma / \dot{\gamma}$, where $\sigma$ is the measured shear stress is monitored under continuous shear at a fixed shear rate $\dot \gamma=$10~s$^{-1}$. The results are summarized in Figure~\ref{fig2}.

In both cases, the normal force is successfully maintained equal to zero (within 2\%) [Fig.~\ref{fig2}(a) and \ref{fig2}(d)], while the gap height decreases by about 4\% over the entire duration of the experiment to compensate for the decrease in the sample volume induced by evaporation [Fig.~\ref{fig2}(b) and \ref{fig2}(e)]. Moreover, the viscosity is measured to be constant throughout both types of experiments, i.e., the \textit{dynamic viscosity} $\eta'$ determined under oscillations, and the \textit{shear viscosity} $\eta$ are both constant in time and within error bars compatible with tabulated viscosity values of water at 25$^{\circ}$C ($\eta$ = 0.89~mPa.s -- see for instance IAPWS 2008) [Fig.~\ref{fig2}(c) and \ref{fig2}(f)]. Note that the viscosity measurements under oscillatory shear [Figs.~\ref{fig2}(c)] show more noise than under continuous shear [Figs.~\ref{fig2}(f)]. This observation can be explained by the fact that the torque applied during an oscillatory shear experiment (2~nN.m) is closer to the lower torque detection limit ($\sim$ 0.5~nN.m) than the torque applied during a steady-shear experiment (420~nN.m). 

Following the same ZNF protocol, we were able to monitor the viscosity of water under continuous shear for more than 10h during which the gap decreases by about 60\% (see Fig.~\ref{figS1} in Appendix~\ref{Appendix1}). The continuous relative gap decrease $\Delta h/h_0$ over time is a direct measurement of the evaporation rate, which allows us to compute the mass flux of water $J_m$ leaving the shear cell as $J_m\simeq 0.3$~g.m$^{-2}$.s$^{-1}$ at 25$^{\circ}$C. This estimate is in good agreement with the values reported in the literature measured with different experimental techniques \cite{Gatapova2014, Kuznetsov2019}. 
Repeating the drying experiment under continuous shear at different temperatures allows us to confirm that the temperature dependence of the evaporation rate follows a scaling that is in agreement with previous work \cite{Sobac:2012} (see Fig.~\ref{figS2} in Appendix~\ref{Appendix2}). 
Furthermore, the viscosity was found to decrease with increasing temperature likewise in agreement with reported values of water viscosity \cite{Cheng:2008}. In conclusion, the results from this section on single-component Newtonian fluids demonstrate the validity of the ZNF protocol.

\subsection{Measuring the viscosity of drying water-glycerol mixtures}
\label{waterglycerol}
In this section, we monitor the viscosity of a Newtonian fluid composed of two miscible liquids, only one of which is volatile. Specifically, we apply the ZNF protocol to the study of drying water-glycerol mixtures by measuring the increase in viscosity as a function of water evaporation. Similar to section~\ref{water}, we monitor the drying of the samples by applying either an oscillatory shear ($\gamma_0=$50\%, $f_0=0.3$~Hz) or a continuous shear ($\dot \gamma=$10~s$^{-1}$), thereby providing measurements of the dynamic viscosity $\eta'$ and the shear viscosity $\eta$, respectively. 

\begin{figure}[t!]
\includegraphics[scale=0.44]{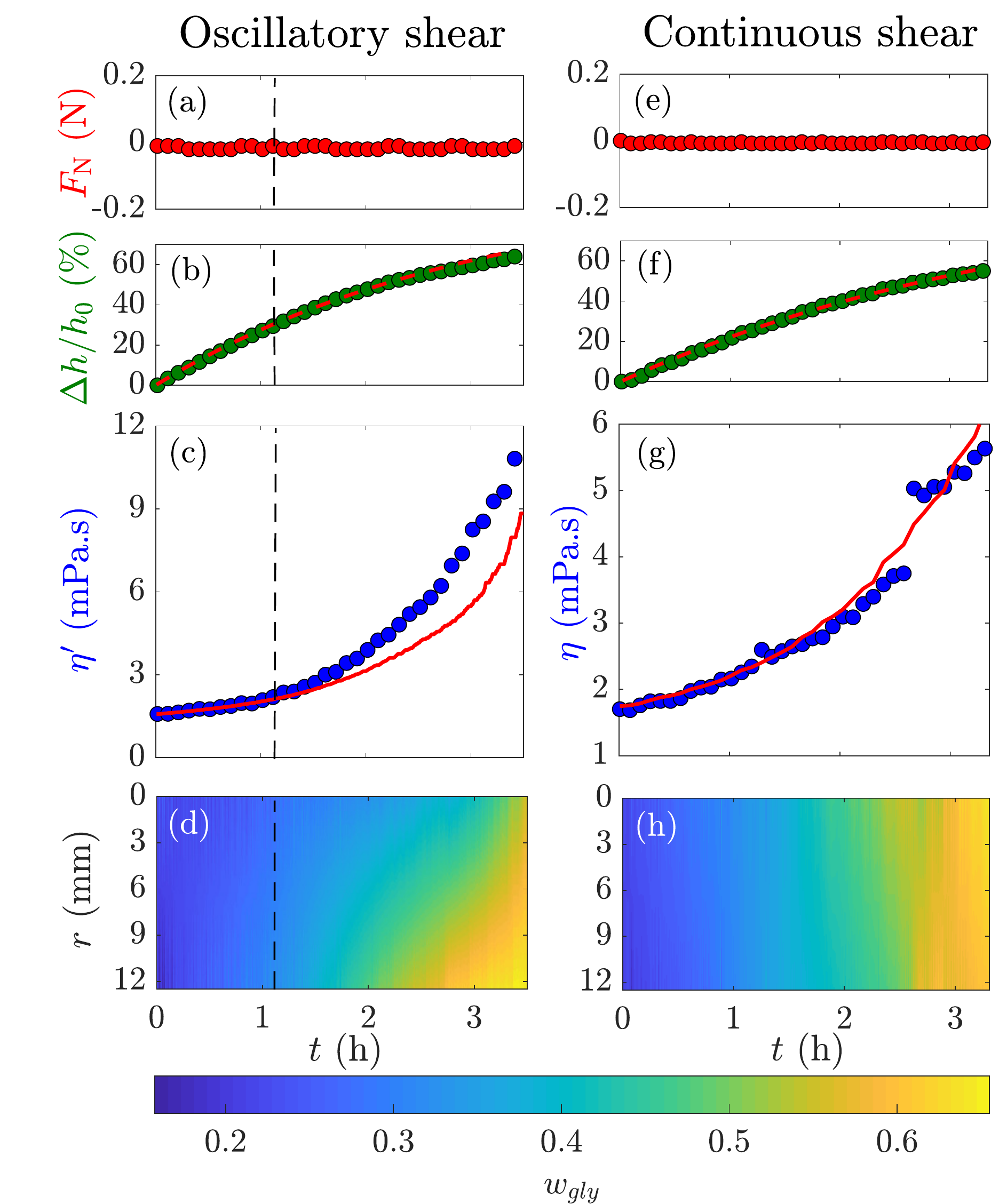}
\caption{\label{fig3} Drying experiments on a water-glycerol mixture with initial glycerol content of 20\% wt. during which the sample's rheological properties are measured with the ZNF protocol under (a--d) oscillatory shear [frequency $f_0=\omega_0/(2\pi)=0.3$~Hz and constant strain amplitude $\gamma_0=50\%$] and (e--h) continuous shear ($\dot{\gamma}=$10 s$^{-1}$). In both cases, we report the temporal evolution of (a,e) the applied normal force $F_{\rm N}$, (b,f) the relative gap decrease $\Delta h/h_0$, (c) the dynamic viscosity $\eta'=G''/\omega_0$, and (g) the shear viscosity $\eta=\sigma / \dot{\gamma}$, and (d,h) a spatio-temporal diagram of an estimated concentration gradient in the sample along the radial direction. Experiments are performed at $T=25^{\circ}$C. In (b) and (f), the dashed red curve corresponds to an exponential fit of the relative gap decrease with a characteristic time of 3.1~h, and 4~h, respectively. In (c) and (g), the continuous red curve corresponds to the tabulated values \cite{Cheng:2008} of the shear viscosity of water-glycerol mixtures at $T=25^{\circ}$C, assuming that the glycerol concentration, inferred from the gap decrease, remains homogeneous. The vertical dashed line in (a-d) corresponds to the time $t=t^*$ beyond which the glycerol concentration is no longer homogeneous along the radial direction.}
\end{figure}

We first discuss results from a mixture with a 20\% wt.~glycerol content, which are reported in Figure~\ref{fig3}. A constant normal force is maintained during 4 hours of both oscillatory and continuous shear [Fig.~\ref{fig3}(a) and \ref{fig3}(e)]. The gap thickness decreases by about 70\% wt.~in both experiments due to water evaporation [Fig.~\ref{fig3}(b) and \ref{fig3}(f)], whereby the weight fraction in glycerol increases as manifested by an increase in sample viscosity [Fig.~\ref{fig3}(c) and \ref{fig3}(g)]. 
Moreover, the relative gap decrease follows an exponential behavior with a characteristic time $\tau \simeq 4.0$~h and 3.1~h for continuous and oscillatory shear tests, respectively. These timescales are shorter than the ones reported for evaporation experiments conducted between static circular plates separated by a fixed gap \cite{Daubersies2011}, most likely due to the presence of additional convective flows in the air surrounding the shear cell, which speed up the evaporation.

Let us now have a closer look at the evolution of the sample dynamic viscosity $\eta'$ reported in Fig.~\ref{fig3}(c). In the first stage of the drying experiment, i.e., for $t<t^*=$ 1.2~h, the dynamic viscosity of the water-glycerol mixture agrees well with tabulated viscosity values of water-glycerol mixtures \cite{Cheng:2008} [continuous red line in Fig.~\ref{fig3}(c)], based on glycerol concentrations determined from the gap decrease [Fig.~\ref{fig3}(b)]. This result suggests that the glycerol concentration is homogeneous in the shear cell. 
However, for $t>t^*$, the dynamic viscosity significantly differs from the tabulated values. This discrepancy most likely originates from the formation of a concentration gradient along the radial direction of the shear cell. Indeed, concentration gradients were reported in drying experiments performed in geometries of fixed volume \cite{Daubersies2011} for Peclet numbers, $Pe = R^2/(D\tau)>1$, where $R$ is the sample size, $D$ is the diffusion coefficient of glycerol in water and $\tau$ is the characteristic time of evaporation. In that context, $D\tau$ represents the diffusive length-scale in the sample of size $R$. For the experiments reported in Fig.~\ref{fig3}(c), Pe $\simeq 11>1$, which supports the idea that the glycerol concentrates at the rim of the parallel-plate geometry, forming a ring of non-volatile fluid, which limits the water evaporation. 
As a result, the drying of the sample slows down, as evidenced by the decrease in speed at which the gap diminishes [Fig.~\ref{fig3}(b)]. In contrast, the viscosity measured under continuous shear is in excellent agreement at all times with literature values \cite{Cheng:2008}, which suggests that the glycerol concentration remains more homogeneous along the radial direction under continuous shear flow conditions.  

\begin{figure*}[th!]
\includegraphics[scale=0.55]{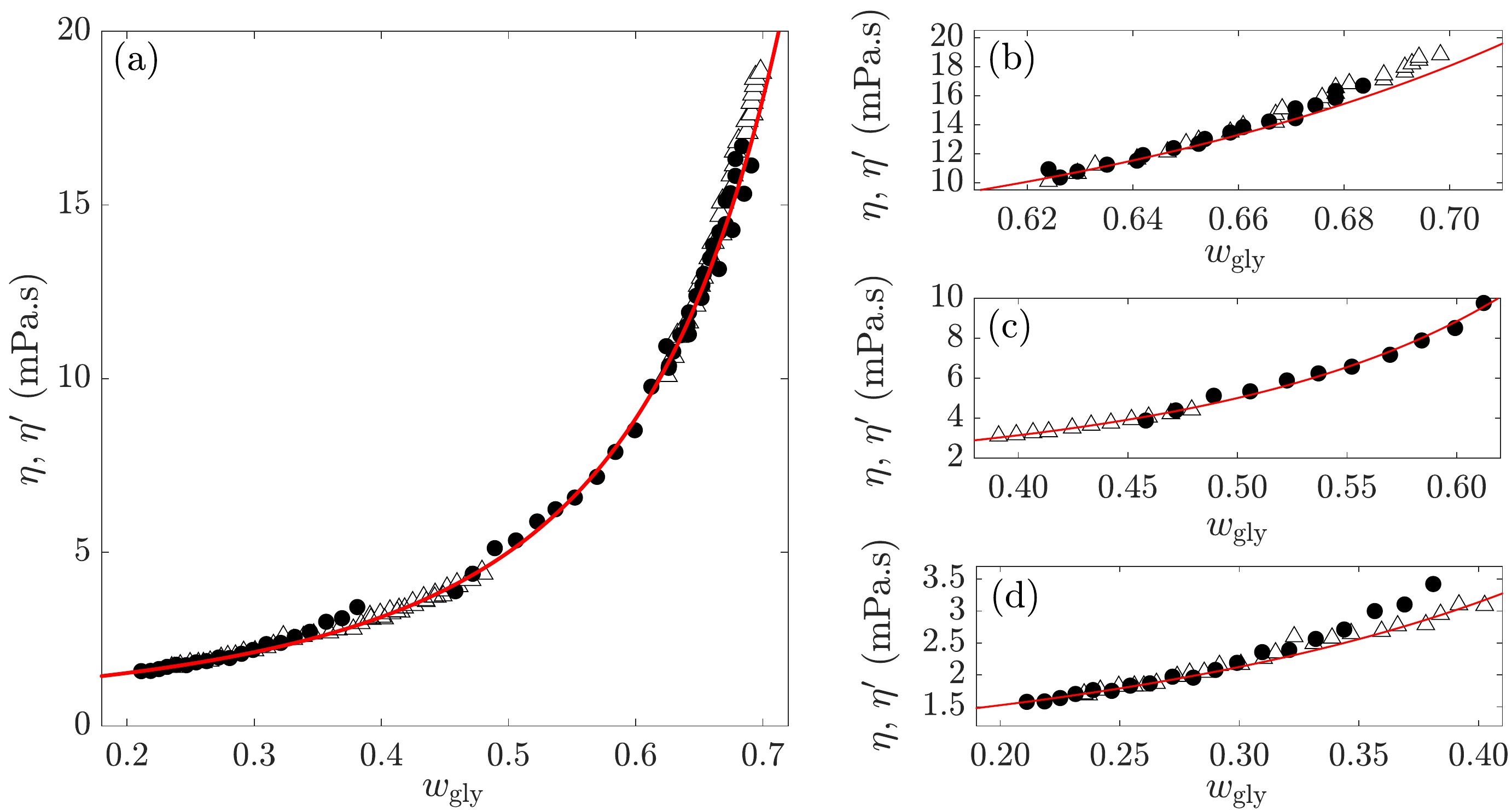}
\caption{\label{fig4} Drying experiments on three different water-glycerol mixtures (initial mass fractions in glycerol of $20$\% wt., 40\% wt.~and 60\% wt.) during which the viscoelastic moduli and the shear viscosity are measured with the ZNF protocol under oscillatory shear and continuous shear, respectively. (a) Dynamic viscosity $\eta'=G''/\omega_0$ ~($\triangle$) and shear viscosity $\eta=\sigma/\dot{\gamma}$ ~($\bullet$). (b--d) Focus on individual experiments with an initial mass fraction of about (b) 60\% wt., (c) 40\% wt., and (d) 20\% wt. Experiments are performed at $T=25^{\circ}$C. Oscillatory measurements are performed at $f_0=0.3$~Hz and $\gamma_0=50$\%, $f_0=0.5$~Hz and $\gamma_0=40$\% and $f_0=0.3$~Hz and $\gamma_0=30$\% for an initial mass fraction of glycerol of 20\% wt., 40\% wt. and 60\% wt. respectively. Continuous shear experiments are performed at $\dot \gamma=10$~s$^{-1}$. The red curve corresponds to the tabulated values of the shear viscosity of water-glycerol mixtures measured at $T=25^{\circ}$C on samples of fixed concentrations, in the absence of any evaporation \cite{Cheng:2008}.}
\end{figure*}

To make sense of the time-evolution of the dynamic viscosity for $t>t^*$ [Figure.~\ref{fig3}(c)], we assume the existence of a linear gradient of glycerol along the radial direction with a larger glycerol concentration at the edge of the plate. The glycerol  concentration at a position $r$ and time $t$ reads: $w_{\rm gly}(r,t)=w_{\rm gly}(0,t) + \alpha(t)r$, where $w_{\rm gly}(0,t)$ is the mass fraction at the center of the geometry ($r=0$) and $\alpha(t)$ the concentration gradient. We can determine the values of $w_{\rm gly}(0,t)$ and $\alpha(t)$ at each point in time, using mass conservation and the viscosity measured with the rheometer during the drying experiment. 
First, the mass conservation in glycerol reads $\bar w_{\rm gly}(t)h(t)=w_{\rm gly}(0,t=0)h(0)$, where $h$ denotes the gap height, $ \bar w_{\rm gly} (t)$ stands for the average mass fraction of glycerol at a time $t$, and $w_{\rm gly}(0,t=0)$ is the initial volume fraction in glycerol, originally homogeneous in the sample. 
The average mass fraction provides a first equation to determine $w_{\rm gly}(0,t)$ and $\alpha(t)$, namely:
\begin{equation}
\bar w_{\rm gly} (t)=\frac{1}{R}\int_{0}
^{R}w_{\rm gly}(r,t){\rm d}r=w_{\rm gly}(0,t)+\alpha(t)\frac{R}{2} \label{eq1}
\end{equation}
Second, the rheometer displays the viscosity of the sample computed at the rim of the geometry, without any single-point correction \cite{Montgomery:2006}. Using tabulated values of the viscosity for water-glycerol mixtures \cite{Cheng:2008}, we compute the mass fraction of glycerol at the edge of the shear cell:
\begin{equation}
w_{\rm gly}(r=R,t)=w_{\rm gly}(0,t)+\alpha(t)R \label{eq2}    
\end{equation} 
Finally, the resolution of the linear system of Eqs.~\eqref{eq1} and~(\ref{eq2}) leads to the determination of the glycerol concentration at the center of the cell $w_{\rm gly}(0,t)$ and the gradient $\alpha(t)$ at all time during the experiment. 

The profile of glycerol concentration obtained with this method is shown as a spatio-temporal diagram in Fig.~\ref{fig3}(d). The same approach applied to the viscosity measurement performed under continuous shear are also reported in Fig.~\ref{fig3}(h). We note that a gradient develops beyond $t^*$ for oscillatory measurements, with $\alpha(t)>0$, whereas the gradient remains negligible $\alpha(t)\simeq 0$ for measurement under continuous shear at $\dot \gamma=10$~s$^{-1}$ (see Fig.~\ref{figS4} in Appendix~\ref{Appendix4}). Continuous shear allows for the homogenization of the water-glycerol mixture while the water evaporates, whereas oscillatory shear is less efficient in doing so. Strikingly, this result for continuous shear depends on the shear intensity $\dot \gamma$, and imposing a larger shear rate leads to a detrimental effect, e.g., imposing $\dot \gamma=100$~s$^{-1}$ leads to the formation of a gradient, with $\alpha(t)<0$ (see Fig.~\ref{figS3} in Appendix~\ref{Appendix3}). In that case, the volume fraction of glycerol is larger at the center of the shear cell as a result of the influence of $\dot \gamma$ on the diffusion of glycerol in water \cite{Rusu:1999}. Therefore, we conclude that a continuous shear of moderate intensity is the most appropriate method to monitor the time-evolution of the viscosity of a dehydrating water-glycerol mixture.     

To illustrate the robustness of our results, we have repeated the drying experiments for water-glycerol mixtures with two other glycerol mass fractions, namely 40\% wt.~and 60\% wt. The viscosity of the drying mixtures is monitored as before, with the ZNF protocol under oscillatory and continuous shear. 
We report the viscosity as a function of the glycerol concentration for the three water-glycerol mixtures (i.e., 20, 40 and 60\% wt.) in Fig.~\ref{fig4}(a), which displays the mixture viscosity as a function of the glycerol weight fraction determined from the gap decrease, assuming that only water is evaporating. The data sets for the three mixtures are reported separately in Fig.~\ref{fig4}(b), (c) and (d).  Except for the long-time measurements in the case of the 20\% wt.~discussed above, all the experimental data are in excellent agreement with tabulated viscosity values of water-glycerol mixtures of various composition, in the absence of any evaporation. 
These results suggests that the 40\%  wt. and 60\%  wt. glycerol solutions are much less prone to develop concentration gradients. Our results therefore demonstrate that the ZNF protocol allows us to accurately monitor the continuously increasing viscosity of a drying Newtonian liquid.

\subsection{Monitoring the drying-induced sol-gel transition in a colloidal suspension}
\label{Ludox}
\begin{figure}
\includegraphics[scale=0.42]{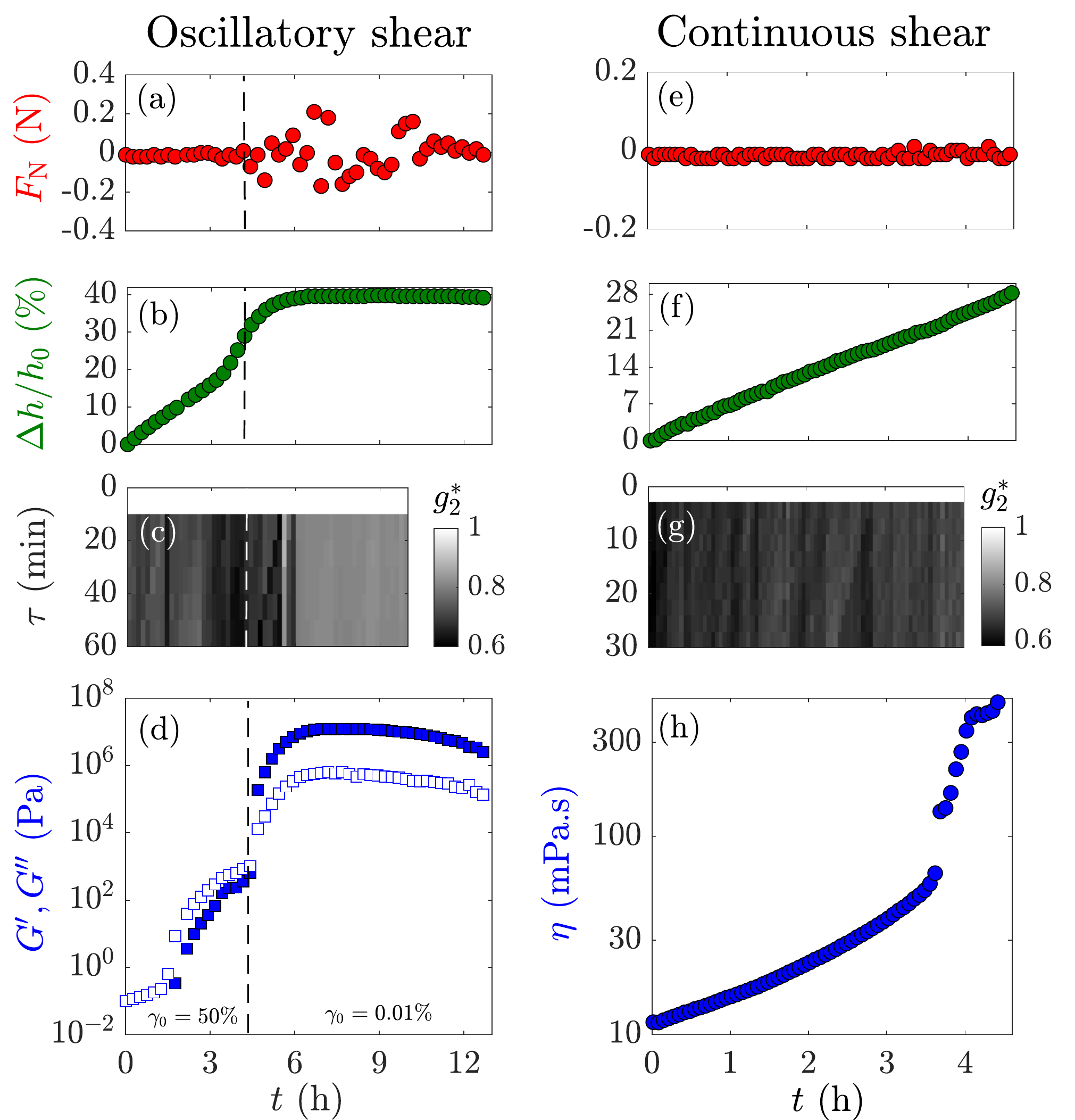}
\caption{\label{fig5} Drying experiments on a suspension of silica nanoparticles of initial volume fraction $\phi_0=0.23$ during which the viscoelastic moduli and the viscosity are measured with the ZNF protocol under (a--d) oscillatory shear [frequency $f_0=1$~Hz, and strain amplitude $\gamma_0=50\%$ up to $t=4.5$~h, and $\gamma_0=0.01\%$ for $t>4.5$~h] and (e--h) continuous shear ($\dot{\gamma}=10$~s$^{-1}$), respectively. In both cases, we report the temporal evolution of (a,e) the applied normal force $F_{\rm N}$, (b,f) the relative gap decrease $\Delta h/h_0$, (c,g) lag-time temporal diagrams computed from the correlation of sample images taken through the transparent bottom plate, (d) the elastic and viscous moduli $G'$ ($\blacksquare$) and $G''$ ($\square$), and (h) the shear viscosity $\eta=\sigma / \dot{\gamma}$.  The vertical dashed line in (d) marks the crossing of $G'$ and $G''$ at $t=t_{\rm c}$. Experiments performed at constant temperature ($T \simeq$21$^{\circ}$C) in a parallel-plate geometry.}
\end{figure}

In this section, we use the ZNF protocol to monitor the rheological properties of a drying aqueous suspension of charged silica nanoparticles. The experiments are here performed in a parallel-plate geometry, whose bottom plate is transparent in order to visualize the sample (see section~\ref{Rheosetup} for details). 

\begin{figure}[t!]
\includegraphics[scale=0.37]{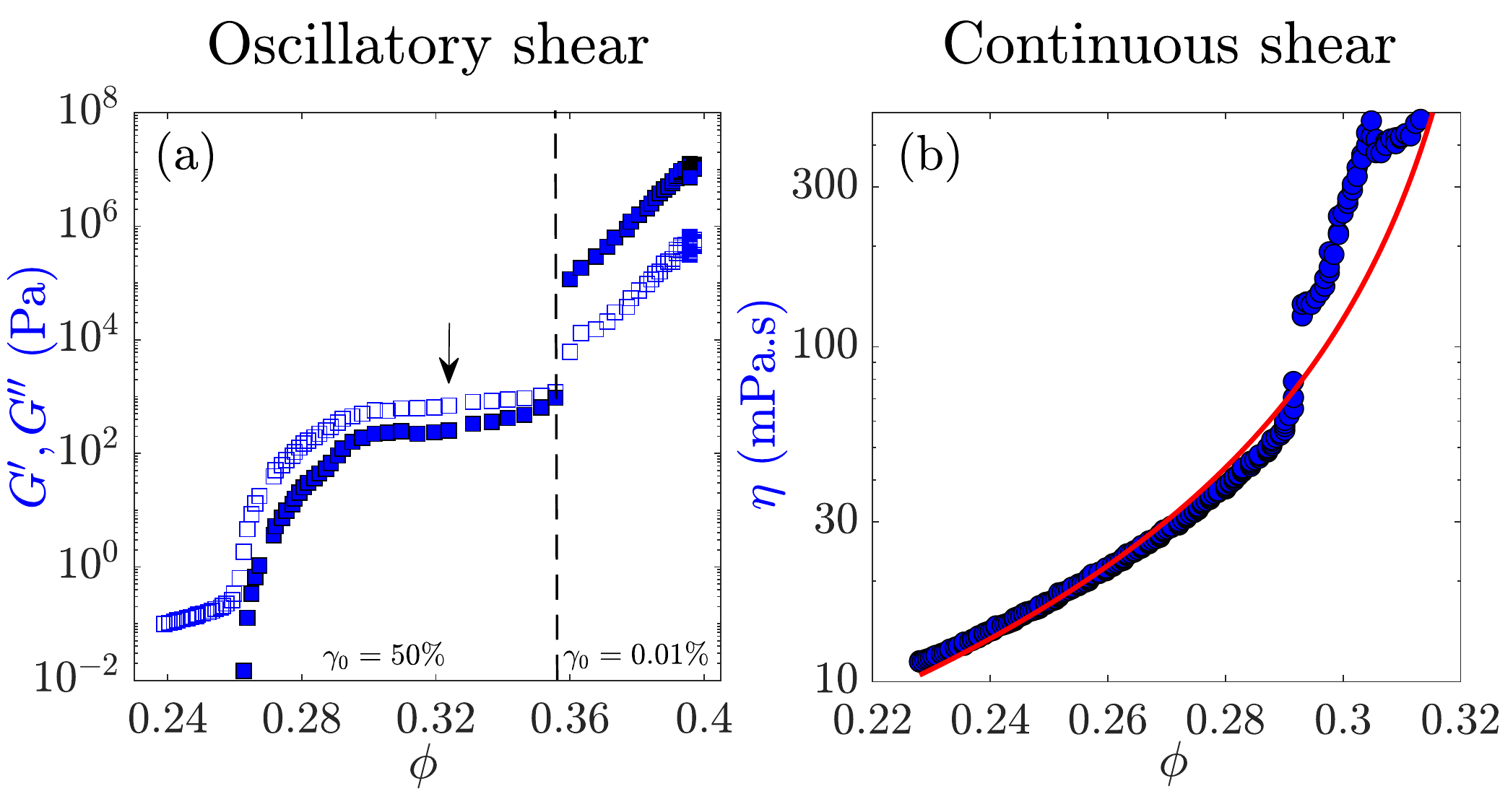}
\caption{\label{fig6} Drying experiment on a suspension of silica nanoparticles of initial volume fraction $\phi_0=0.23$. Same data as in Fig.~\ref{fig5}(d) and \ref{fig5}(h) reported as a function of the volume fraction $\phi$ inferred from the gap decrease [Fig.~\ref{fig5}(f)]. (a)  $G'$ ($\blacksquare$) and $G''$ ($\square$) vs $\phi$. The vertical dashed line highlights the volume fraction at which the strain amplitude is switched from $\gamma_0=50$\% to $\gamma_0=0.01$\%. The black arrow highlights the critical $\phi^{({\rm ref})}_c=0.32$ beyond which a yield stress is expected \cite{DiGiuseppe:2012}. (b) Shear viscosity $\eta=\sigma/\dot{\gamma}$ vs $\phi$. The red curve corresponds to the Quemada model [see Eq.~\eqref{eq:Quemada} in the text] with $\phi_c=0.33$. Experiments performed at constant temperature ($T \simeq$21$^{\circ}$C) in a parallel-plate geometry, with a transparent bottom plate.} 
\end{figure}

We first discuss the case of a drying aqueous suspension monitored by oscillatory shear. The linear viscoelastic moduli are determined at $f_0=1$~Hz, with a variable strain amplitude $\gamma_0$, which is adjusted to the sample viscoelastic properties so as to optimize the sensitivity of the measurement, without affecting the gelation process \cite{Ewoldt:2015}. We found empirically that a good compromise consists in imposing a large strain amplitude $\gamma_0=50$\% while the sample is mainly viscous before switching to a small amplitude $\gamma_0=0.01$\% as soon as $G'$ becomes comparable to $G''$ to avoid damaging the soft solid that is forming as a result of the solvent evaporation.
Such a strain-adapted protocol allows us to monitor the temporal evolution of the elastic and viscous modulus of the suspension over 12h, while the solvent evaporates. The results are plotted as a function of time in Fig.~\ref{fig5}(d), and the same data are reported in Fig.~\ref{fig6}(a) as a function of the volume fraction, which is inferred from the gap variation.

At the start of the experiment, the sample is liquid and characterized solely by its viscous modulus $G''$, which slowly increases as the solvent evaporates [Fig.~\ref{fig5}(d)]. At $t\simeq 1.5$~h, the sample shows a linear viscoelastic response such that $G'<G''$. Beyond $t\simeq 1.5$~h, both modulus increase as a result of the solvent evaporation up to $t \simeq 4.5$~h at which $G'$ crosses $G''$, marking the emergence of a yield stress. The crossing of $G'$ and $G''$ marked by a dashed vertical line in Fig.~\ref{fig5}(d) and Fig.~\ref{fig6}(a) occurs at a volume fraction $\phi_c\simeq 0.35$, which is in reasonable agreement with the value expected for Ludox HS-40, i.e., $\phi^{({\rm ref})}_c \simeq 0.32$ \cite{Loussert:2016, DiGiuseppe:2012, Boulogne:2014, Ziane:2015}. Note that up to $\phi_c$, the normal force is correctly maintained within the interval $F_{\rm N}=(0.00\pm$0.01)~N [Fig.~\ref{fig5}(a)], while the relative gap decrease speeds up before reaching $\phi_c$ [Fig.~\ref{fig5}(b)]. Beyond the intersection of $G'$ an $G''$, i.e., for $t>4.5$~h (or $\phi>0.35$), both $G'$ and $G''$ increase up to $t \simeq 6$~h, where they reach a plateau such that $G'>G''$. Interestingly, the viscoelastic plateau coincides with the cessation of the gap decrease, which suggests that the evaporation has completely stopped beyond $t=6$~h. Finally, for $t>9$~h, both $G'$ and $G''$ decrease while the gap stays constant, which hints at some local reorganization of the colloidal glass, without any supplemental loss of solvent.      

To confirm this scenario, we take advantage of the transparent bottom plate of the shear cell in which the drying experiment is conducted (see section~\ref{Rheosetup}). Images of the entire surface of the sample shot through the transparent bottom plate and taken every 10~min allow us to quantify the local dynamics associated with the evaporation. The degree of correlation between two images separated by a lag time $\tau$ are determined by the ensemble-average intensity correlation function $g_2(t,\tau)$ defined as follows:
\begin{equation} \label{eq:G2}
g_2(t,\tau)=\frac{\langle I_p(t).I_p(t+\tau)\rangle_p}{\langle I_p(t) \rangle_p.\langle I_p(t+\tau) \rangle_p}
\end{equation}
where $I_p$ represents the brightness of the pixel $p$, $\langle ... \rangle_p$ the average over all the pixels \cite{Cipelletti:2002,Divoux:2015}. The correlation function is further normalized into a function noted $g_2^*(t,\tau)$, which satisfies $g_2^*(t,\tau=0)=1$.
The normalized correlation function is computed at a frame rate of about 1.6~mHz with a lag time $\tau$ ranging between 0 and 1~h. The results are pictured in a lag-time temporal diagram in Fig.~\ref{fig5}(c). At any point in time for $t<6$~h, the images slowly de-correlates over  $\tau=1$~h. The rate of decorrelation decreases abruptly in the vicinity of the gelation point ($t_c=4.5$~h) before stopping completely at $t=6$~h when the elastic and viscous modulus reach a plateau. The lack of decorrelation beyond $6$~h confirms that the cessation of the gap decrease visible in Fig.~\ref{fig5}(b) marks the end of the evaporation, which results from the formation of a solid crust at the sample edge. Such a solid crusts isolating the sample from the ambient atmosphere is indeed visible, when separating the two plates at the end of an experiment.     

We now turn to drying experiments performed on the suspension of silica particles under continuous shear, at $\dot \gamma=10$~s$^{-1}$. The results, including the shear viscosity, are reported as a function of time in Fig.~\ref{fig5}(e--h), and of the volume fraction in Fig.~\ref{fig6}(b). As the solvent evaporates, the viscosity rises over time at increasing speed [Fig.~\ref{fig5}(h)]. The normal force is correctly set to $F_{\rm N}=(0.00\pm$0.01)~N [Fig.~\ref{fig5}(e)], while the relative gap decrease shows a somewhat linear trend over the 4.5~h that the experiment lasts [Fig.~\ref{fig5}(f)]. Note that the total relative gap decrease is  $\Delta h/h_0 \simeq 28$\% at $t=4.5$~h, which is comparable to the value reached at the same time for experiments performed under oscillatory shear [see Fig.~\ref{fig5}(b)]. However, here the gap decrease accompanies a spatially homogeneous increase of the particle volume fraction, as confirmed by images of the drying suspension. Indeed, the normalized correlation function $g_2^*(t,\tau)$ pictured in Fig.~\ref{fig5}(g) shows roughly the same rate of decorrelation as a function of the lag-time $\tau$ for the entire duration of the experiment. This observation shows that, in stark contrast with the experiments performed under oscillatory shear, drying experiments performed under constant shear yield a spatially more homogeneous sample, thus preventing the formation of a solid crust.  

As supplemental evidence that the drying colloidal suspensions remains homogeneous under continuous shear, we find that the corresponding viscosity measured as a function of time, and reported in Fig.~\ref{fig6}(b) as a function of the volume fraction $\phi$, is well-described by the Quemada model that reads \cite{Quemada:1977}:
\begin{equation}
\label{eq:Quemada}
\eta = \eta_s \left(1-\frac{\phi}{\phi_{\rm max}}\right)^{-2}
\end{equation}
where $\eta_s$ is the viscosity of the solvent and $\phi_{\rm max}$ is the volume fraction at maximum packing. Such a model has been successfully applied to describe the rheology of Ludox suspensions \cite{DiGiuseppe:2012,Sobac:2020}. 
Here too, this phenomenological model describes our data very well with $\phi_{\rm max} = 0.33$, in good agreement with the value expected for suspensions for Ludox HS-40. The fact that $\phi_{\rm max}<\phi_c$, and the good agreement of $\phi_{\rm max}$ with the literature, confirm that the sample remains more spatially homogeneous during the experiment performed under continuous shear. Indeed, assuming a linear gradient of concentration following the steps discussed in section~\ref{waterglycerol} yields negligible values of $\alpha$ [see Fig.~\ref{figS4}(b) in Appendix~\ref{Appendix4}].
This result shows that drying experiments performed under continuous shear yield more accurate values of $\phi_c$ than experiments performed under oscillatory shear.  
  
\section{Conclusion}
The ZNF protocol has been successfully applied to perform time-resolved rheological measurements during the drying of Newtonian fluids and colloidal suspensions. This protocol can be employed with both oscillatory shear and continuous shear to measure in a single experiment with only one sample loading what is usually obtained with multiple measurements on samples of various concentrations, in the absence of evaporation. Moreover, we have shown that in comparison to oscillatory shear experiments, continuous shear at moderate intensity allows for the homogenization of the sample, which limits the growth of spatial gradients and therefore leads to more accurate rheological measurements. This work constitutes a proof of concept and opens new perspectives for determining the macroscopic rheological properties of viscoelastic samples subject to solvent evaporation, and monitor their eventual sol-gel or glass transition. Moreover, we have observed that depending on the shear intensity applied during the dehydration of the sample, one may control the spatial gradient of concentration that builds up in a sample, which is sandwiched between two parallel disks and dries along the radial direction. As such, beyond providing an accurate method for bulk rheological measurements on drying samples, our work suggests a simple way for making solids with spatially modulated mechanical properties. Future work will be focused on the characterization of the local mechanical properties of such heterogeneous solids produced from the drying of colloidal suspensions under external shear. 

\begin{acknowledgments}
The authors kindly acknowledge M. Lenz, S. Manneville, B. Mao, G.H. McKinley, J.-B. Salmon and P. Snabre for fruitful discussions, and the support from the MIT-France program from the MIT International Science and Technology Initiative. 
\end{acknowledgments}

\appendix 
\section{long-duration drying experiment on water}
\label{Appendix1}
\begin{figure}[!h]
\includegraphics[scale=0.40]{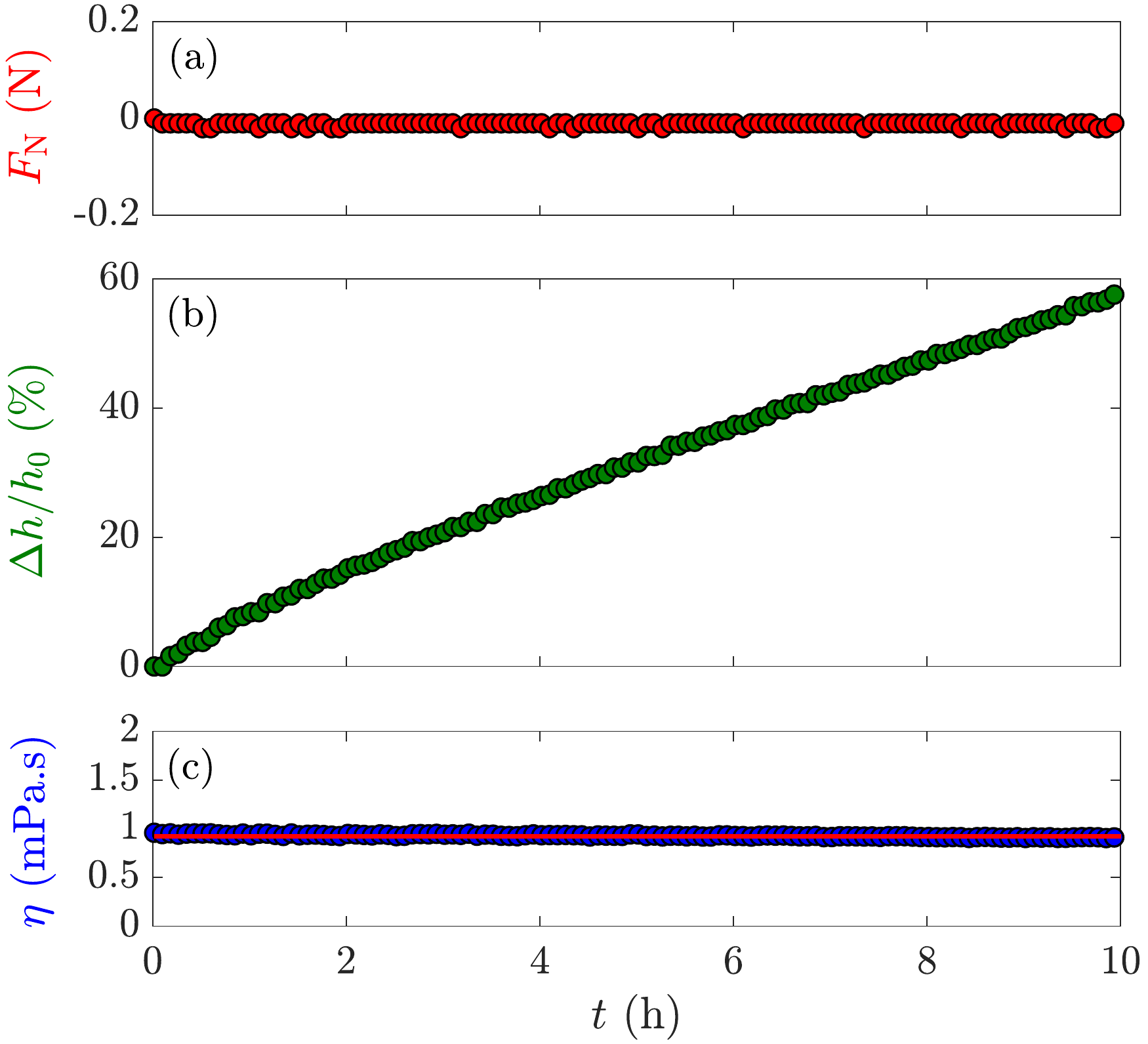}
\caption{\label{figS1} Viscosity measurement under continuous shear ($\dot{\gamma}=$10 s$^{-1}$) on a drying water sample using the ZNF protocol over 10 hours. Temporal evolution of (a) the applied normal force $F_{\rm N}$, (b) the relative gap decrease $\Delta h/h_0$, and (c) the shear viscosity $\eta=\sigma / \dot{\gamma}$. Experiments performed at $T=25^{\circ}$C, with $h_0=500$~$\mu$m. In (c), the red line corresponds to the mean value of the viscosity computed over the whole duration of the experiment $\eta=(0.92 \pm 0.02$)~mPa.s.}
\end{figure}

To complement the drying experiments performed on distilled water over a short duration of typically 30~min, we report here the case of a 10h-long experiment. The results, displayed in Fig.~\ref{figS1}, show that the ZNF protocol allows us to measure accurately the viscosity of water, while the volume of the sample decreases by 60\%.  
\section{Impact of temperature on the drying rate}
\label{Appendix2}
\begin{figure}[!t]
\includegraphics[scale=0.42]{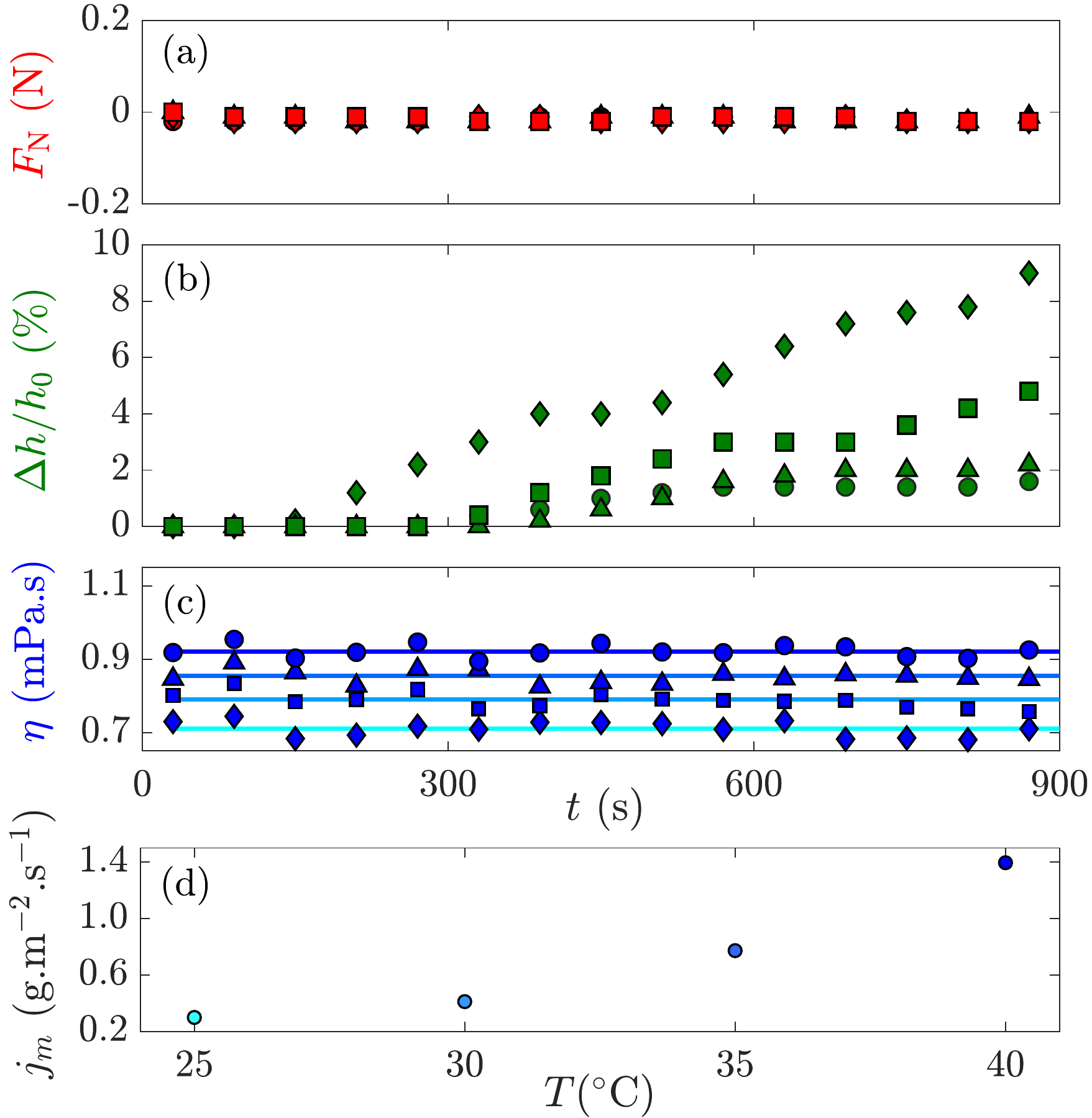}
\caption{\label{figS2} Viscosity measurement with the ZNF protocol under continuous shear ($\dot{\gamma}=$10 s$^{-1}$) on a drying water sample at different temperatures. Temporal evolution of (a) the applied normal force $F_{\rm N}$, (b) the relative gap decrease $\Delta h/h_0$, and (c) the shear viscosity $\eta=\sigma / \dot{\gamma}$. Experiments performed at $T=25^{\circ}$C ($\triangle$), 30$^{\circ}$C ($\circ$), 35$^{\circ}$C ($\square$) and 40$^{\circ}$C ($\diamond$), with $h_0=500$~$\mu$m. In (c), the continuous lines correspond to the mean value of the viscosity computed over the duration of the experiment, i.e., $\eta=(0.92 \pm 0.01)$~mPa.s, $\eta=(0.85\pm 0.02)$~mPa.s, $\eta=(0.78\pm 0.02)$~mPa.s, and  $\eta=(0.71 \pm 0.02)$~mPa.s for increasing temperature. (d) Mass fluxes of water leaving the shear cell due to evaporation vs temperature.}
\end{figure}

Here we show in Fig.~\ref{figS2} that the ZNF protocol allows to monitor accurately the viscosity of distilled water drying at different temperatures, ranging between 25$^{\circ}$C and 40$^{\circ}$C.

\section{Estimate of the linear gradient} \label{Appendix4}
\begin{figure}[!bh]
\includegraphics[scale=0.4]{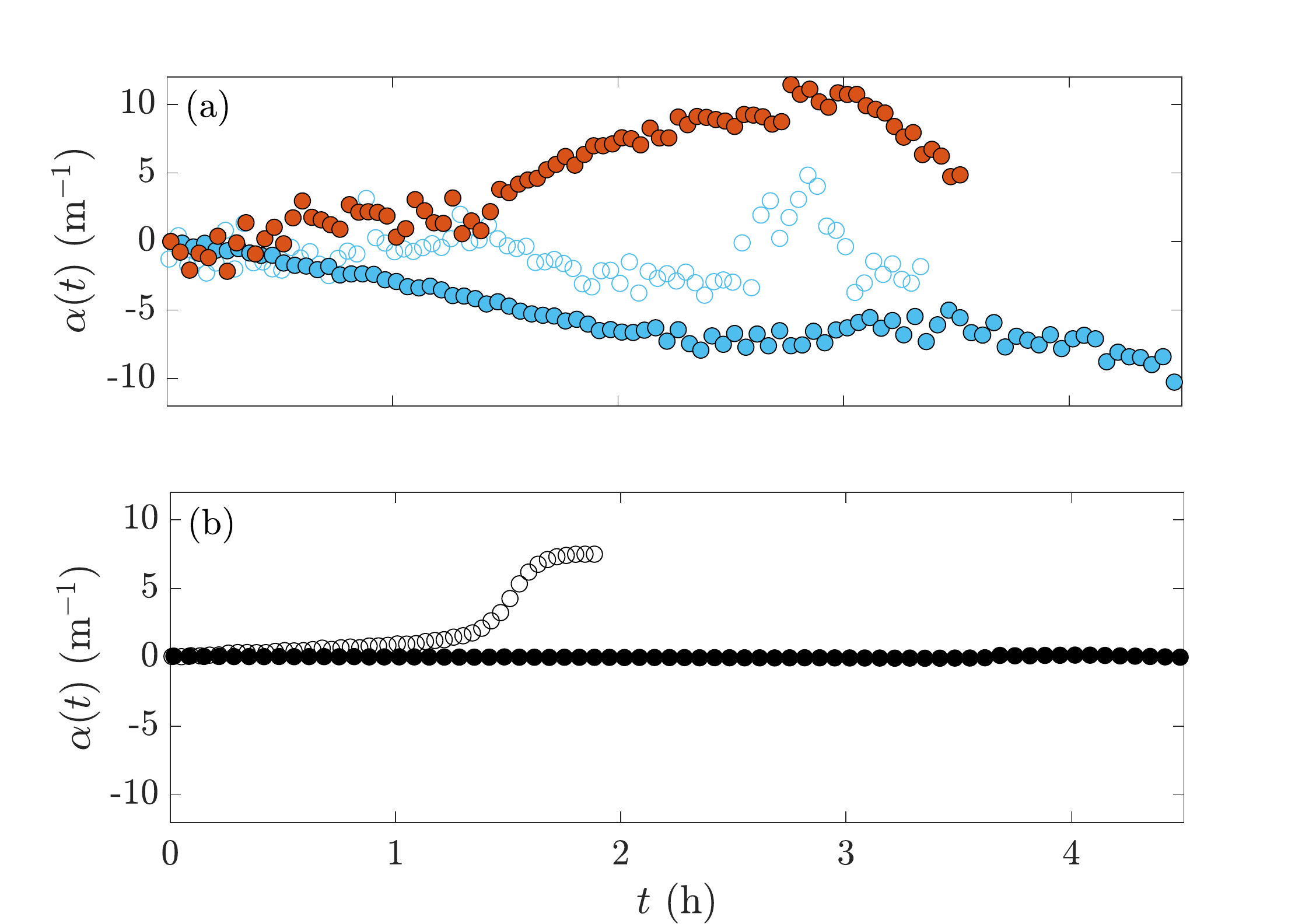}
\caption{\label{figS4} Temporal evolution of slopes of the linear gradients $\alpha$ for (a) a water-glycerol mixture with an initial mass fraction of 20\% wt. under oscillatory shear ($f_0 = 0.5$~Hz, $\gamma_0 = 50\% $: (\textcolor{orange}{$\bullet$}), and continuous shear [$\dot{\gamma}=10$~s$^{-1}$ : (\textcolor{cyan}{$\circ$}), $\dot{\gamma}=100$~s$^{-1}$ : (\textcolor{cyan}{$\bullet$})], and for (b) a suspension of silica nanoparticles of initial volume fraction $\phi_0 = 0.23$ under oscillatory shear before the crossing of $G'$ and $G''$ [$f_0 = 1$~Hz, $\gamma_0 = 50\%$ : ($\circ$)] and continuous shear [$\dot{\gamma} = 10$~s$^{-1}$: ($\bullet$)].}
\end{figure}

We report in Fig.~\ref{figS4} the values of the model linear gradients $\alpha$ for the experiments performed on water-glycerol mixtures [Fig.~\ref{figS4}(a)] and suspensions of silica colloids [Fig.~\ref{figS4}(b)]. We show that the concentration gradients due to the solvent evaporation are positive under oscillatory shear and continuous shear of moderate intensity ($\dot \gamma=10$~s$^{-1}$), whereas the gradient is negative under continuous shear of larger intensity ($\dot \gamma=100$~s$^{-1}$). We do not observe any gradient for a drying experiment performed  on a suspension of silica particles under continuous shear ($\dot \gamma=100$~s$^{-1}$).  

\section{Drying water-glycerol mixture under large continuous shear} \label{Appendix3}
\begin{figure}[!th]
\includegraphics[scale=0.45]{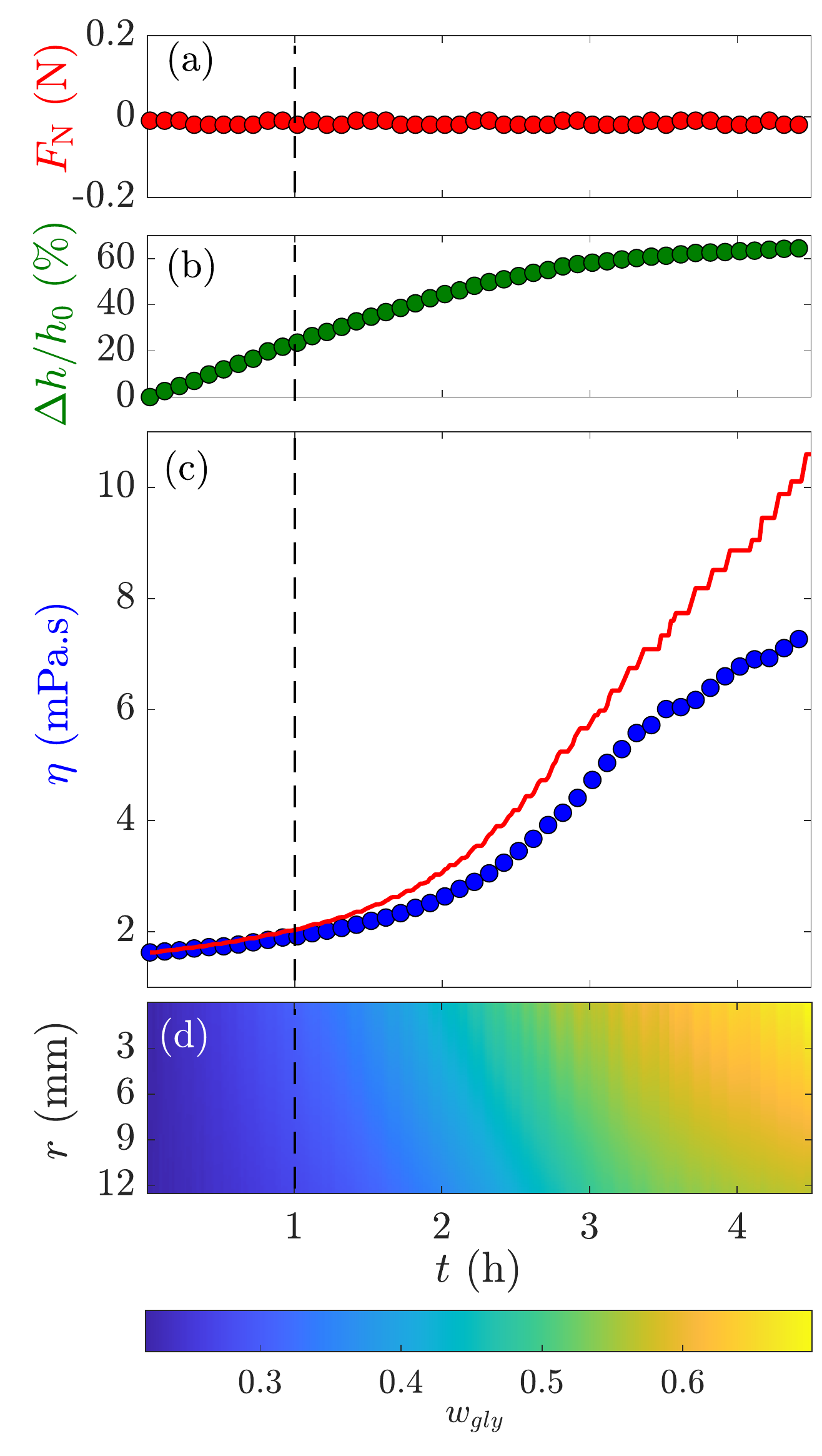}
\caption{\label{figS3} Viscosity measurement with the ZNF protocol under continuous shear ($\dot{\gamma}=100$ ~s$^{-1}$) on a drying water-glycerol mixture, with an initial glycerol content of 20\% wt. Temporal evolution of (a) the applied normal force $F_{\rm N}$, (b) the relative gap decrease $\Delta h/h_0$, (c) the shear viscosity $\eta=\sigma/\dot \gamma$, and (d) the computed temporal evolution of a linear gradient in the sample. Experiment is performed at $T=25^{\circ}$C. The continuous red curve corresponds to the tabulated values \cite{Cheng:2008} of the shear viscosity of water-glycerol mixtures at $T=25^{\circ}$C, assuming that the glycerol concentration, inferred from the gap decrease, remains homogeneous at all time. The vertical vertical dashed line corresponds to the time beyond which the glycerol concentration is not homogeneous along the radial direction.}
\end{figure}

We show in Fig.\ref{figS3} the viscosity measurement of a water glycerol mixture (initial content of 20\% wt. in glycerol) under continuous shear ($\dot{\gamma}=100$~s$^{-1}$). Contrary to the case of $\dot{\gamma}=10$~s$^{-1}$ that is reported in the main text in Fig.~\ref{fig3}(e-h), one can see the formation of a positive gradient of concentration, i.e.,  water accumulates at the periphery of the shear cell. After $t=1$~h, one cane see the departure of the measured viscosity from the expected value for spatially homogeneous drying.  


%

\end{document}